\documentclass[sigconf]{acmart}

\usepackage{latexsym}
\usepackage{booktabs} 
\usepackage{multirow}
\usepackage[normalem]{ulem}
\usepackage{enumitem}
\usepackage{subcaption}
\usepackage{algorithm}
\usepackage{algorithmic}
\usepackage{tabularx}
\newcommand{\etal}{{\emph{et al.}}}
\AtBeginDocument{%
  }
\usepackage{multirow}

\setcopyright{acmlicensed}
\copyrightyear{2025}
\acmYear{2025}
\acmConference[MM '25]{Proceedings of the 33th ACM International Conference on Multimedia}{October 27--31, 2025}{Dublin, Ireland}
\acmBooktitle{Proceedings of the 33th ACM International Conference on Multimedia (MM '25), October 27--31, 2025, Dublin, Ireland}
\acmDOI{XXXXXXX.XXXXXXX}

\acmSubmissionID{3743}

\begin{document}
\title{
KEN: Knowledge Augmentation and Emotion Guidance Network for Multimodal Fake News Detection}

\author{Peican Zhu}
\authornote{Peican Zhu and Yubo Jing are joint first authors.}
\affiliation{%
  \institution{Northwestern Polytechnical University}
  \city{Xi’an}
  \country{China}
}
\email{ericcan@nwpu.edu.cn}

\author{Yubo Jing}
\authornotemark[1] 
\affiliation{%
  \institution{Northwestern Polytechnical University}
  \city{Xi’an}
  \country{China}
}
\email{jingyubo@mail.nwpu.edu.cn}

\author{Le Cheng}
\affiliation{%
  \institution{Northwestern Polytechnical University}
  \city{Xi’an}
  \country{China}
}
\email{chengle@mail.nwpu.edu.cn}

\author{Keke Tang}
\authornote{Keke Tang and Yangming Guo are joint corresponding authors.}
\affiliation{%
 \institution{Guangzhou University}
 \city{Guangzhou}
  \country{China}
 }
\email{tangbohutbh@gmail.com}

\author{Yangming Guo}
\authornotemark[2]
\affiliation{%
  \institution{Northwestern Polytechnical University}
  \city{Xi’an}
  \country{China}
  }
\email{yangming_g@nwpu.edu.cn}


\begin{abstract}
In recent years, the rampant spread of misinformation on social media has made accurate detection of multimodal fake news a critical research focus.
However, previous research has not adequately understood the semantics of images, and models struggle to discern news authenticity with limited textual information.
Meanwhile, treating all emotional types of news uniformly without tailored approaches further leads to performance degradation.
Therefore, we propose a novel \textbf{K}nowledge Augmentation and \textbf{E}motion Guidance \textbf{N}etwork (KEN).
On the one hand, we effectively leverage LVLM’s powerful semantic understanding and extensive world knowledge.
For images, the generated captions provide a comprehensive understanding of image content and scenes, while for text, the retrieved evidence helps break the information silos caused by the closed and limited text and context. 
On the other hand, we consider inter-class differences between different emotional types of news through balanced learning, achieving fine-grained modeling of the relationship between emotional types and authenticity.
Extensive experiments on two real-world datasets demonstrate the superiority of our KEN.

\end{abstract}



\maketitle

\begin{figure}[t]
\setlength{\abovecaptionskip}{0cm}
\centering
\scalebox{1}
{\includegraphics[width=1\linewidth]{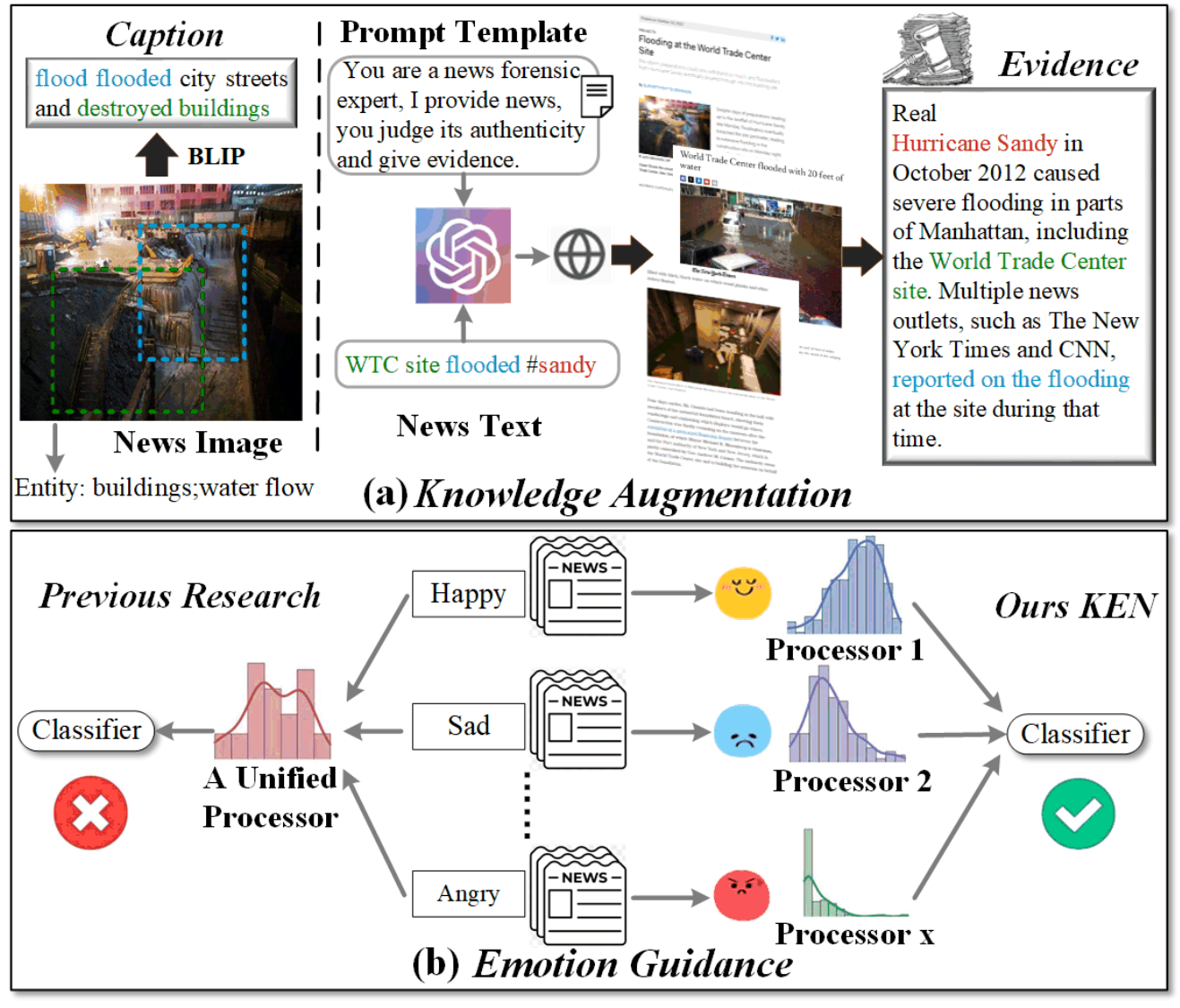}}
\caption{
Advantages of our proposed model.
(a)
The generated caption captures image semantics and scenes, while the evidence breaks the information silos caused by the closed and limited text and context.
(b) Previous studies overlooked inter-class differences among news with different emotions, treating all types of news with a unified processor, failing to leverage their strengths and reducing performance.
}
\label{fig:intro}
\end{figure}

\section{Introduction}
Social media has become an indispensable part of our daily life, which has significantly enhanced the dissemination and exchange of information due to its convenience. 
However, it has also become a fertile ground for the spread and proliferation of fake news~\cite{de2021approaches}. 
Well-designed fake news is indistinguishable, and multimodal formats capture readers' interest.
Meanwhile, social interactions and sharing behaviors further amplify the impact of rumors.
It seriously jeopardizes national security, economic development and social stability~\cite{cheng2025efficient}. 
Therefore, fake news detection become a problem that needs to be solved urgently.

Existing fake news detection methods can be categorized into unimodal and multimodal approaches based on the news content. 
With the diversification of news formats, multimodal approaches have received increasing attention.  They mainly include two perspectives: one is the cross-modal interaction and fusion~\cite{Att-RNN,MCAN,zhu2024multimodal,hua2023multimodal,HMCAN,CMMTN}, and the other is the application of alignment and consistency~\cite{chen2022cross,safe,FEND-CLIP,CCGN}.
The former relies on the extraction, fusion, and enhancement of information from different modalities, involving methods such as pre-trained models, inter-modal interaction, and entity extraction.
The latter tends to learn cross-modal differences and balance the relationship between unimodal and multimodal features.
These methods have achieved excellent detection results.
However, there are still shortcomings in previous works.

\textbf{1) Insufficient news information results in inadequate semantic understanding and poor fusion performance}.
Previous research in the image modality has focused on entity-level and pixel-level information, which is beneficial for detecting image tampering but has not truly understood the semantics and scenes of images. 
In the text modality, limited by local and closed information, it is easy to fall into information silos. 
As shown in Figure~\ref{fig:intro}(a), in this example, the news image is not tampered with, and entities such as ‘buildings’ and ‘water flow’ can be retrieved. 
However, in the news text, only ‘flooded’ weakly corresponds to the image entity ‘water flow’.
The low semantic correlation between the image and text leads to poor fusion.
Moreover, terms like ‘WTC site’ and ‘sandy’ are unclear and perplexing in terms of their meaning.
The isolated context also put detection models in a dilemma.
As a result, the limited content and information silos make it difficult for previous models to comprehend the semantics and prone to misclassification.
To address this issue, we propose a knowledge augmentation method.
First, the generated caption describes the scenes and semantics of the image, with key information like ‘flood flooded’ and ‘destroyed buildings’ helping us form an initial understanding of the news. 
Next, evidence retrieved and analyzed through LVLMs clarifies that ‘sandy’ refers to ‘Hurricane Sandy’, and ‘WTC site’ refers to ‘World Trade Center site’. 
The phrase ‘reported on the flooding’ further confirms the authenticity of the flooding at WTC site caused by hurricane sandy.
Additionally, our model achieves knowledge denoising and fusion, and designs three perspectives to enhance the expression and integration of multimodal semantics.

\textbf{2) Ignoring inter-class differences among news of different emotional types leads to degraded performance}.
The generation of fake news is closely related to the malicious intent of its creators, and the spread of rumors often varies with the emotional tone expressed in news.
Starting with basic emotion classification, we find that fake news in real life tends to express negative emotions, while real news usually appears positive and optimistic.
Therefore, different emotional categories can help in detecting the authenticity of news.
As shown in Figure~\ref{fig:intro}(b), news emotions are categorized into more detailed types.
When the news emotion is happy, the feature distribution of the news tends to be more positive and real, leaning toward the right side. 
However, when the emotion is anger, the news features are more likely to be negative and fake, leaning toward the left side.
In other words, there are differences in feature distribution across different emotional categories of news.
However, previous research has overlooked this phenomenon and applied a unified detector to process all news features, leading to reduced detection performance.
To address this issue, we propose an emotion guidance method.
First, we analyze the emotional expressions in multimodal news.
Then, a balanced learning strategy is designed to map news features into more suitable emotion-specific spaces for specialized learning.
In addition, we introduce an emotion reasoning task to assist in fake news detection.

In this paper, we propose a novel \textbf{K}nowledge Augmentation and \textbf{E}motion Guidance \textbf{N}etwork (KEN) for multimodal fake news detection.
First, this framework effectively leverages the powerful semantic understanding and extensive world knowledge of LVLMs. 
For images, the generated captions fully comprehend the content and context of images, while for text, the obtained evidence helps break the information silos caused by the closed and limited context.
Next, we design three different fusion perspectives, utilizing the co-attention mechanism and CLIP to enhance feature fusion and improve the semantic representation of news. 
Unlike existing detection methods, we consider inter-class differences between different emotional types of news and perform fine-grained modeling of the relationship between emotional types and veracity.
We feed news features into the feature mapping spaces of the corresponding emotional domains processors for learning, which improves overall detection performance and avoids the performance degradation caused by treating all emotional types of news uniformly in traditional methods.
Additionally, we design a coarse-grained emotion reasoning task to assist fake news detection.
KEN is extensively evaluated on two real-world datasets, and the experimental results demonstrate its effectiveness and superiority.

Our contributions are summarized as follows:
\begin{itemize}
    \item We introduce Knowledge Augmentation, which leverages evidence to break the information silos caused by closed and limited text and context while generating captions to describe the content and scenes of images, enabling the model to achieve a more comprehensive semantic understanding and judgment basis for news.
    \item We propose Emotion Guidance, which considers inter-class differences between different emotional types of news through balanced learning and achieves fine-grained modeling of the relationship between emotional types and authenticity, addressing performance degradation caused by traditional methods that uniformly process news of all emotional types.
    \item Extensive experiments show that our proposed KEN consistently outperforms SOTA fake news detection baselines on two competitive datasets.
\end{itemize}

\section{Related Work}

\subsection{Unimodal Fake News Detection}
Early methods relied on manual analysis of news content and language styles~\cite{kumar2016disinformation,ott2011finding}. However, they required a significant amount of time and effort. 
With the development of deep learning, current approaches are mainly divided into context-based and content-based methods. Context-based methods leverage the dissemination structure of news and user interaction information~\cite{yin2024gamc,cheng2024gin}. 
However, they face practical challenges, as their reliance on extensive foundational data or graph propagation structures introduces significant complexity in data collection and processing, and also makes them vulnerable to perturbation-based attacks that undermine graph-based reasoning~\cite{zhu2024general,he2025hypergraph}.
Content-based methods are further divided into two types, i.e., unimodal and multimodal. Some works used unimodal methods based on CNN~\cite{CNN1} and RNN~\cite{RNN1} to learn the content, style, and stance-based features of text for detecting fake news. 
MVNN~\cite{MVNN} considered learning both the frequency and pixel domain features of visual content to identify fake news. 
However, unimodal methods have failed to employ the interaction and fusion of multimodal information, leading to unsatisfactory detection performance.

\subsection{Multimodal Fake News Detection}
Multimodal fake news detection methods have received increasing attention from scholars. 
Att-RNN~\cite{Att-RNN} introduced social context features for knowledge expansion. 
Spotfake~\cite{Spotfake} proposed to integrate pre-trained BERT~\cite{BERT} and VGG-19~\cite{VGG} to extract features. EANN~\cite{EANN} designed the event detection task to obtain event-specific features to improve the detection performance. 
MVAE~\cite{MVAE} introduced a variational autoencoder to reconstruct data and employed adversarial learning to enhance feature representation. 
Some studies~\cite{EM-FEND,KAN} proposed methods for enhancing and aggregating features through visual and textual entities. KMGCN~\cite{KMGCN} obtained semantically enhanced features by acquiring external knowledge highly relevant to news entities from the real world. 
HMCAN~\cite{HMCAN} and MCAN~\cite{MCAN} employed co-attention~\cite{co-attention} layers to enhance the interaction and fusion of inter-modal features.
MRHFR~\cite{MRHFR} proposed a cognition-aware fusion method inspired by human reading habits, and inferred comment-news consistency through a coherence constraint reasoning layer.
However, the aforementioned methods fail to leverage the powerful cross-modal learning and knowledge analysis capabilities of large models, resulting in insufficient semantic understanding and poor fusion performance.
With the emergence of large multimodal model CLIP~\cite{CLIP}, CAFE~\cite{chen2022cross} drew inspiration from its consistency evaluation and proposed cross-modal ambiguity to balance the influence of unimodal and multimodal features. 
Some works~\cite{FEND-CLIP,MMFN} used the reinforcement capabilities of CLIP to further learn the semantic consistency across different modalities. 
However, many real news also suffer from the issue of mismatch between visual and textual content, so relying solely on consistency is unreliable.
The widespread dissemination of fake news is also closely tied to the emotional reactions it evokes in readers, and emotional features are sometimes used to aid in the detection of fake news. Giachanou~\etal~\cite{giachanou2019leveraging} embedded emotional signals as features and combined them with textual content to detect fake news. 
Ghanem~\etal~\cite{ghanem2021fakeflow} believed that emotional factors could manipulate readers' emotions, and therefore used deep learning to evaluate the emotional information flow in fake news, combining the extracted emotional data with news topics to detect fake news. 
Both Zhang~\etal~\cite{zhang2021mining} and Luvembe~\etal~\cite{luvembe2023dual} explored the relationship between the emotions of the publisher and the audience, assisting in fake news detection from the perspective of emotional resonance. However,  these methods do not fully explore the application of emotions in fake news detection.

Existing research has been effective in extracting multimodal features from news content and achieving relatively good performance. However, there are still some shortcomings:
1) These methods focus on pixel-level information in images but fail to deeply understand the image semantics and scenes. In text, they can only learn localized and closed information, leading to the problem of information silos. These issues result in insufficient semantic understanding and poor fusion performance.
2) They overlook inter-class differences between different emotional types of news, feeding all news into a unified fake news detector without specializing the processing based on emotional domains, thereby degrading detection performance.
Therefore, we propose the \textbf{K}nowledge augmentation and \textbf{E}motion guidance \textbf{N}etwork (KEN) to address these issues.

\section{Methodology}
\begin{figure*}[t]
\centering
\setlength{\abovecaptionskip}{0cm}
\scalebox{0.98}{\includegraphics[width=18cm]{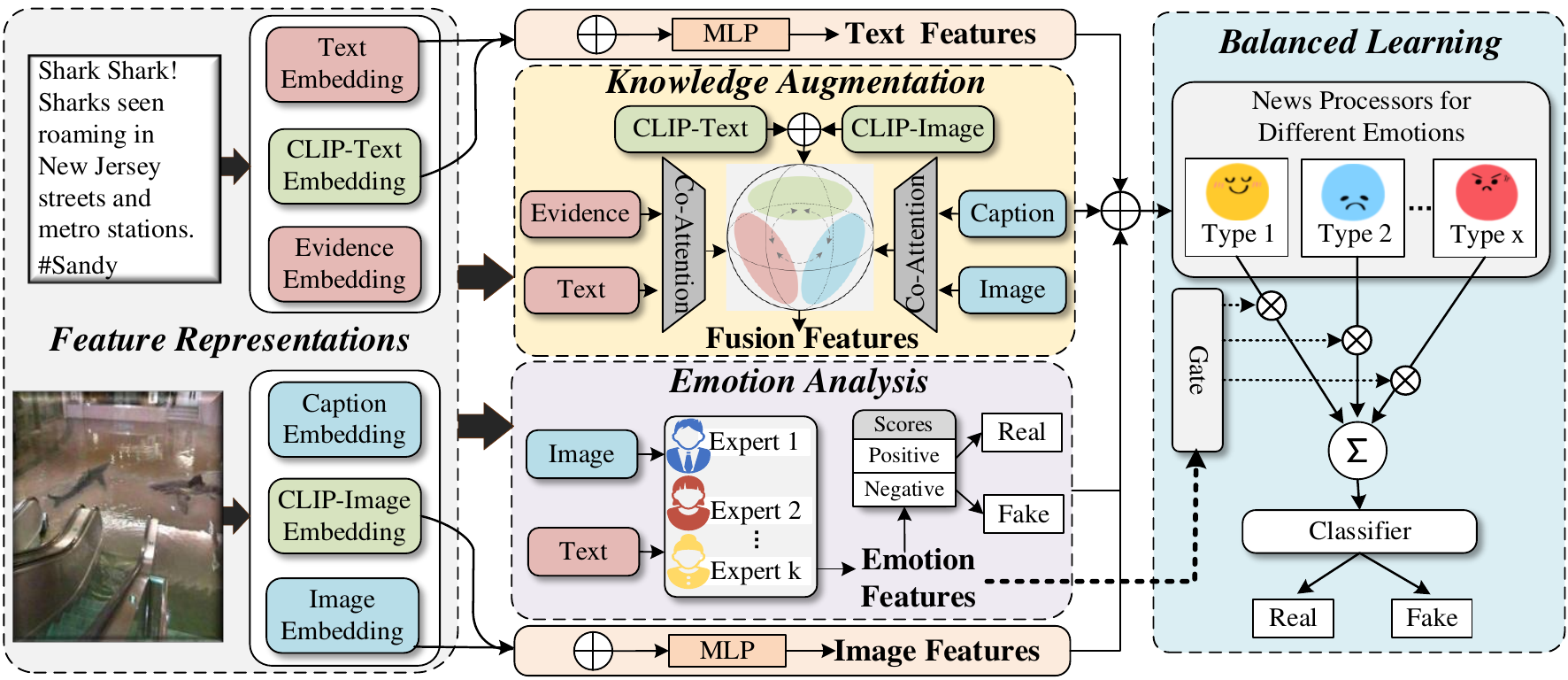}}
\caption{Illustration of the proposed KEN. It contains of four components: Feature Representations, Knowledge Augmentation, Emotion Analysis and Balanced Learning.}
\label{fig:method}
\end{figure*}

The comprehensive architecture of KEN is illustrated in Figure~\ref{fig:method}.
First, we encode the original textual and visual content features.
Second, we generate captions to describe and understand the scenes and semantics of images, while obtaining evidence to break information silos caused by limited text and context.
Then, the co-attention mechanism facilitates semantic complementarity and enhancement between captions and images, as well as between evidence and text.
Using pre-trained CLIP, we obtain highly aligned image-text features to guide fusion feature acquisition and unimodal enhancement.
Third, we analyze the emotions of images and text to obtain the overall emotional features, associating the emotional tendency of news with its authenticity in a coarse-grained manner.
Finally, in balanced learning, we employ emotional features and a gating mechanism to highlight processors corresponding to the emotional types of news in a fine-grained manner. 
The final aggregated features are fed into a classifier for authenticity prediction.

\subsection{Feature Representations}

\noindent{\textbf{Image and Text Embedding.}}
Social media news typically consists of images and text.
We denote social media news as $N$, where included images and text are represented as $V$ and $T$, respectively.
For images $V$, we use Swin-T~\cite{Swin-T} to learn local semantic relationships, denoted as $V=\{v_1,v_2,...,v_n\}$, where $n$ is the number of regions in each image.
For text $T$, we use BERT~\cite{BERT} to encode and learn semantic relationships between words, denoted as $T=\{t_1,t_2,...,t_m\}$, where $m$ is the length of the text.

\noindent{\textbf{Knowledge Acquisition.}}
As shown in Figure~\ref{fig:intro}(a), we obtain captions and evidence for images and text, respectively.
Through joint pre-training of vision and language modalities, BLIP2~\cite{BLIP} significantly enhances image understanding, offering powerful cross-modal representations for tasks like visual question answering.
We utilize it to generate captions for images and encode them, denoted as 
$P=\{p_1,p_2,...,p_z\}$, where $z$ is the length of the captions.
For text, we design a prompt template and utilize GPT-4 along with web search to obtain evidence related to the news. 
This provides preliminary reasoning and analysis, supplements information, enriches the scenes, and mitigates the information silo effect caused by limited content and context.
We use BERT to encode the evidence, denoted as $R=\{r_1,r_2,...,r_u\}$, where $u$ is the length of the evidence.

\noindent{\textbf{CLIP Enhanced Embedding.}}
The CLIP model~\cite{CLIP} maps semantically similar images and text into adjacent semantic spaces through an image-text matching task.
Thus, we use the pre-trained CLIP to obtain enhanced image embeddings $v_c$ and text embeddings $t_c$.

\subsection{Knowledge Augmentation}
The obtained evidence $R$ and captions $P$ not only provide rich semantic and contextual information but also introduce some noise.
Therefore, we employ two co-attention blocks to achieve interactive fusion and denoising between text $T$ and evidence $R$, as well as between images $V$ and captions $P$, each consisting of two Transformer~\cite{BERT} encoders.
The following is illustrated with text $T$ and evidence $R$ as inputs to a Transformer encoder.

Firstly, we obtain the attention distribution over the evidence $R$ by performing a dot product operation between the evidence $R$ and the text $T$.
Then, based on the attention weights, the parts of the evidence that are highly relevant to the text in terms of content and semantics are highlighted, while less relevant noisy information is suppressed.
To capture key semantic information in the evidence more comprehensively, we employ a multihead attention mechanism, which is processed simultaneously by $i$ parallel computing units.
On this basis, important semantic information in the evidence $R$ is extracted to supplement and enhance the text $T$.
Finally, we obtain the evidence-enhanced text representation $h_{r\to t}$:

\begin{equation}
\begin{gathered}
h=softmax(\frac{(TW_q)\cdot(RW_k)^T}{\sqrt{d_h}})\cdot(RW_v)\\
MultiHead(T,R,R)=[h_1;h_2;...;h_i]\cdot W_i \\
h^\prime=Norm(T+MultiHead(T,R,R))\\
h_{r\to t}=Norm(h'+\sigma(h'))
\end{gathered}
\end{equation}
where $T$, $R$, $R$ are query, key, and value matrix, respectively. 
$W_q$, $W_k$, $W_v$ are mapping matrices used to project the information into the same space.
$h$ is the semantic information highly related to the text, obtained from the evidence by a parallel unit.
The symbol ; denotes vector concatenation and $W_i$ is a trainable parameter.
$Norm$ and $\sigma$ are the normalization method and feed forward network.

The text-enhanced evidence representation $h_{t\to r}$ can also be learned within the same co-attentive block. 
Then, $h_{r\to t}$ and $h_{t\to r}$ are aggregated to obtain the integrated content and semantic representation $\mathcal S_1$ for the text-evidence perspective.
Similarly, another co-attention block obtains the caption-enhanced image representation $h_{p\to v}$ and the image-enhanced caption representation $h_{v\to p}$, leading to the integrated content and semantic representation $\mathcal S_2$ for the image-caption perspective.
Due to the content and semantic gap between features of the above two perspectives,
we introduce CLIP-enhanced multimodal features $\mathcal S_3$ to improve cross-modal similarity and strengthen semantic associations between the two modalities, thereby enhancing cross-modal feature fusion.
The features from the three different perspectives are represented as follows:

\begin{equation}
\begin{gathered}
\mathcal S_1=\sigma_1\left(concat(h_{r\to t},h_{t\to r})\right)\\
\mathcal S_2=\sigma_2\left(concat(h_{p\to v},h_{v\to p})\right)\\
\mathcal S_3=\sigma_3(concat(t_c,v_c))\\
\end{gathered}
\end{equation}
where $\sigma_i$ denotes the feed forward network  that maps the features to the same semantic space.
Finally, the features from the three perspectives undergo cross-modal interaction learning in the same semantic space to obtain the initial fusion features $\mathcal{M}_{f}$.
Additionally, considering that low-matching cross-modal features may degrade the quality of fusion features, we introduce the text-image similarity $\theta=\cos(t_c,v_c)$ to measure the correlation. When the correlation between modalities is strong, the fusion features can be prominently expressed. Conversely, when the correlation is weak, the model prioritizes preserving unimodal features 
to mitigate the impact of ineffective or erroneous modal interactions.

\begin{equation}
\mathcal{M}_{f}=\theta\cdot\sigma_f(concat(\mathcal S_1,\mathcal S_2,\mathcal S_3))\\
\end{equation}

For unimodal features, we leverage the outputs of the CLIP model for semantic enhancement and optimization, improving representation while preserving the original feature space structure.
Ultimately, this not only enhances the semantic expressiveness of each unimodal feature but also leverages strong cross-modal alignment, providing a solid foundation for obtaining more powerful aggregated features.
The final unimodal text features $\mathcal{M}_{t}$ and image features $\mathcal{M}_{v}$ can be expressed as:

\begin{equation}
\begin{gathered}
\mathcal{M}_{t}=\sigma_t(concat(t_c,T))\\
\mathcal{M}_{v}=\sigma_v(concat(v_c,V))
\end{gathered}
\end{equation}

\subsection{Emotion Analysis}
We input the text $T$ and image $V$ into the emotion analysis to obtain the overall emotion features of news.
To prevent arbitrary evaluation by a single model, we employ $k$ experts to comprehensively analyze the emotion features for each modality.
Taking the structure of the $k$-th expert as an example, we first use Bi-LSTM to capture emotion semantics within text sequences and image regions, respectively.
Then, the self-attention mechanism highlights the  more important emotion expressions in each modality.
Subsequently, the feed forward network extracts the emotion features of text and image, denoted as $e_t^k$ and $e_v^k$.
Finally, we perform an averaging operation on the $k$ experts to obtain the aggregated textual emotion features $e_t$ and visual emotion features $e_v$.
The emotion analysis using image $V$ as an example can be represented as follows:

\begin{equation}
\begin{gathered}
e_v^k=\sigma(SelfAtt(BiLSTM(V)))\\
e_v=AVG(e_v^1,e_v^2,...,e_v^k)
\end{gathered}
\end{equation}

Considering the varying emotional significance expressed by different modalities, we introduce a modulation parameter $\gamma$ to balance the influence of image and text emotions when analyzing the overall emotion features of the news.
It is worth noting that the parameter $\gamma$ ranges between 0 and 1. A larger value indicates that the emotional information expressed by the text modality is more crucial, while a smaller value suggests that the emotional information expressed by the image modality is more prominent.
Finally, the textual emotion features $e_t$ and visual emotion features $e_v$ are adjusted by the parameter $\gamma$ and passed through the feed forward network to obtain the final news emotion feature $\mathcal{M}_{e}$:
\begin{equation}
\mathcal{M}_{e}=\sigma_e(concat(\gamma e_t,(1-\gamma)e_v))
\end{equation}

Considering that fake news tends to exhibit negative emotions, while real news tends to express positive emotions, we coarsely construct an auxiliary task. We perform a binary classification task on the emotional features of the news, $\mathcal{M}_{e}$, to distinguish between positive and negative emotions. Furthermore, we assume that news with negative emotions is fake, while news with positive emotions is real, thereby constructing the emotion reasoning loss $\mathcal{L}_{emo}$:
\begin{equation}
\begin{gathered}
\hat{y}^{e} = softmax(\mathcal{M}_{e}W_e+b_e)\\
\mathcal{L}_{emo}=\sum-[ylog(\hat{y}^{e})+(1-y)log(1-\hat{y}^{e})]
\end{gathered}
\end{equation}
where $y^e$ represents the predicted binary emotion classification result of the news, and $y$ denotes the authenticity label of news.

\subsection{Balanced Learning}

We acquire all necessary features for classification, including text features($\mathcal{M}_{t}$), image features ($\mathcal{M}_{v}$), emotion features  ($\mathcal{M}_{e}$), and fusion features ($\mathcal{M}_{f}$).
These features are concatenated to form a comprehensive representation of news that integrates both content and emotions,
i.e.,
$\mathcal{M}=\begin{bmatrix}\mathcal{M}_{t};\mathcal{M}_{v};\mathcal{M}_{e};\mathcal{M}_{f}\end{bmatrix}$.
Next, we consider inter-class differences between different emotional types of news and map the aggregated news features into the relevant emotion domain spaces for learning, avoiding the performance degradation caused by traditional methods that uniformly process all emotional types of news.

First, we feed the aggregated news features $\mathcal{M}$ to $x$ different processors $H_x$ to learn the feature distribution patterns of news content and authenticity within the current emotional domain:
\begin{equation}
m_x=H_x(\mathcal{M};\theta_x)\\
\end{equation}
where $m_x$ represents the news features learned in different emotional domain processors, and $\theta_x$ denotes the learnable parameters within processors.
Meanwhile, we feed the emotional features $\mathcal{M}_{e}$ into the gating mechanism to identify the primary emotional type expressed in the news.
Subsequently, based on the relevance weights between the news and different emotional types, we highlight  the feature mapping spaces of the corresponding emotional domains processors.
Therefore, we obtain the aggregated news features $\mathcal{F}$ for authenticity classification:

\begin{equation}
\begin{gathered}
\boldsymbol{a}=[a_1,a_2,...,a_x]=softmax\left(G(\mathcal{M}_{e};\theta_g)\right)\\
\mathcal{F}=\sum a_xm_x\\
\end{gathered}
\end{equation}

Finally, we pass the news feature $\mathcal{F}$ through a fully connected layer to determine the authenticity of the news and ultimately obtain the predicted label $\hat{y}^{f}$.
Then, the rumor classification loss, denoted as $\mathcal{L}_{fnd}$, is formulated using cross-entropy loss as follows:
\begin{equation}
\begin{gathered}
\hat{y}^{f}=softmax(\mathcal{F}W_f+b_f)\\
\mathcal{L}_{fnd}=\sum-[ylog(\hat{y}^{f})+(1-y)log(1-\hat{y}^{f})]
\end{gathered}
\end{equation}
where $y$ indicates the ground truth label of the news.
The overall loss $\mathcal{L}$ for our KEN consists of rumor classification loss $\mathcal{L}_{fnd}$ and emotion reasoning loss $\mathcal{L}_{emo}$:
\begin{equation}
\mathcal{L}=\mathcal{L}_{fnd}+\lambda\mathcal{L}_{emo}
\end{equation}
where $\lambda$ is a loss adjustment parameter.

\section{Experiments}

\subsection{Experimental Setup}
\subsubsection{Dataset Description.}
We conduct experiments on two real-world datasets, i.e., Weibo~\cite{Att-RNN} and Twitter~\cite{twitter}.
In this work, we primarily focus on textual and visual information.
Weibo contains 3643 real news and 4203 fake news with 9528 images.
Twitter contains 8720 real news and 7448 fake news with 514 images.
We follow the same pre-processing steps and data splits as the benchmark on the two datasets.
To evaluate the performance, we adopt accuracy, precision, recall, and F1-score as metrics.

\subsubsection{Baselines.}
For comprehensive comparison, we consider several state-of-the-art baselines: 

$\bullet$ \textbf{Att\_RNN} \cite{Att-RNN} 
uses LSTM to extract textual features and VGG-19 to extract visual features, and employs an attention mechanism to fuse textual, visual, and social context information.

$\bullet$ \textbf{EANN} \cite{EANN} is a multi-task learning framework based on event detection. It introduces an event discriminator component, which derives event-invariant features through adversarial networks. 

$\bullet$ \textbf{SAFE} \cite{safe} uses a pre-trained model to convert image features into text semantics and assists fake news detection by measuring the similarity between the image and text features.

$\bullet$ \textbf{CAFE} \cite{chen2022cross} leverages BERT and ResNet to learn text and image features, respectively. By measuring cross-modal ambiguity and adaptively aggregating features, it addresses misclassification issues caused by discrepancies between different modalities.

$\bullet$ \textbf{MKEMN} \cite{MKEMN} retrieves relevant concepts and event knowledge through entity information, then uses convolution operations to fuse text, image, and knowledge information. Finally, the integrated features obtained from the fusion are used for fake news detection.

$\bullet$ \textbf{EMAF} \cite{EMAF} extracts entity information from images and texts separately, compares and aligns entity information from different modalities for learning, and improves detection performance through fine-grained interactive learning with images and texts based on comprehensive entity features.

$\bullet$ \textbf{MCAN} \cite{MCAN} extracts spatial and frequency domain features of images through VGG-19 and CNNs respectively, and uses BERT to learn text features. By continuously stacking the co-attention layer for deep fusion interaction. 

$\bullet$ \textbf{HMCAN} \cite{HMCAN} is a hierarchical multimodal contextual attention network that combines multimodal contextual information and the hierarchical semantics of text through co-attention. It deeply learns the inter- and intra-modality relationships, significantly improving detection accuracy.

$\bullet$ \textbf{CMMTN} \cite{CMMTN} uses a multimodal masked transformer to align text and image features, while masking irrelevant context between modalities. It employs BERT and ResNet for feature extraction and a curriculum-based PU learning method to handle positive and unlabeled data for fake news detection.

$\bullet$ \textbf{CCGN} \cite{CCGN} uses vision GNN to suppress irrelevant image background information and consistency contrastive learning to minimize the semantic distance between text and image features, with cross-attention for improved fusion.

\subsubsection{Implementation Details.}
All experiments are conducted on NVIDIA RTX A6000 GPUs with 48GB memory, and the proposed KEN is implemented using the PyTorch framework. 
For image encoding, we use the swin-base-patch4-window7-224 model, setting the input image size to 224 × 224, with the number of regions ($n$) in each image fixed at 49. 
We employ the bert-base-chinese model for the Weibo dataset and the bert-base-uncased model for the Twitter dataset. 
For the text, captions, and evidence, we set their lengths $m$, $z$, and $u$ to 300, with embedding dimensions of 768 for all.
We use the pre-trained CLIP model is ViT-B/16.
For the co-attention and self-attention mechanisms, the number of heads ($i$) is set to 8, and a dropout rate of 0.5 is applied. 
The number of experts performing emotion analysis ($k$) is set to 3, while the number of emotional types ($x$) is set to 5.
For the emotion adjustment parameter $\gamma$ and the loss adjustment parameter $\lambda$, we set $\gamma$ and $\lambda$ to 0.7 and 0.2 for Weibo, and to 0.3 and 0.75 for Twitter, respectively.
The model is trained for 40 epochs using the Adam optimizer with a learning rate of 0.001, and a batch size of 16.

\begin{table*}
\setlength{\abovecaptionskip}{0cm} 
\centering
\caption{Experimental results of baselines and our KEN on Weibo and Twitter datasets.}
\begin{center}
\resizebox{0.75\textwidth}{!}{
\begin{tabular}{ccccccccc}
\hline
\multirow{2}{*}{Dataset} & \multirow{2}{*}{Methods} & \multirow{2}{*}{Accuracy} & \multicolumn{3}{c}{Fake News} & \multicolumn{3}{c}{Real News} \\ \cline{4-9} 
 &  &  & Precision & Recall & F1-score & Precision & Recall & F1-score \\ \hline
\multirow{11}{*}{Weibo} 
 & Att\_RNN~\cite{Att-RNN} & 0.772 & 0.854 & 0.656 & 0.742 & 0.720 & 0.889 & 0.795 \\
 & EANN~\cite{EANN} & 0.782 & 0.827 & 0.697 & 0.756 & 0.752 & 0.863 & 0.804 \\
 & SAFE~\cite{safe} & 0.816 & 0.818 & 0.815 & 0.817 & 0.816 & 0.818 & 0.817 \\
 & CAFE~\cite{chen2022cross} & 0.840 & 0.855 & 0.830 & 0.842 & 0.825 & 0.851 & 0.837 \\
 & MKEMN~\cite{MKEMN} & 0.814 & 0.823 & 0.799 & 0.812 & 0.723 & 0.819 & 0.798 \\
 & EMAF~\cite{EMAF} & 0.874 & 0.895 & 0.857 & 0.876 & 0.862 & 0.897 & 0.879 \\
 & MCAN~\cite{MCAN} & 0.899 & 0.913 & 0.889 & 0.901 & 0.884 & 0.909 & 0.897 \\
 & HMCAN~\cite{HMCAN} & 0.885 & 0.920 & 0.845 & 0.881 & 0.856 & 0.926 & 0.890 \\
 & CMMTN~\cite{CMMTN} & 0.889 & 0.886 & 0.893 & 0.889 & 0.892 & 0.885 & 0.893 \\
 & CCGN~\cite{CCGN} & 0.908 & 0.922 & 0.894 & 0.908 & 0.894 & 0.922 & 0.908 \\
 & \textbf{KEN} & \textbf{0.935} & \textbf{0.937} & \textbf{0.934} & \textbf{0.935} & \textbf{0.932} & \textbf{0.935} & \textbf{0.934} 
 \\ \hline
\multirow{10}{*}{Twitter} & Att\_RNN~\cite{Att-RNN} & 0.664 & 0.749 & 0.615 & 0.676 & 0.589 & 0.728 & 0.651 \\
 & EANN~\cite{EANN} & 0.648 & 0.810 & 0.498 & 0.617 & 0.584 & 0.759 & 0.660 \\
 & SAFE~\cite{safe} & 0.762 & 0.831 & 0.724 & 0.774 & 0.695 & 0.811 & 0.748 \\
 & CAFE~\cite{chen2022cross} & 0.806 & 0.807 & 0.799 & 0.803 & 0.805 & 0.813 & 0.809 \\
 & MKEMN~\cite{MKEMN} & 0.715 & 0.814 & 0.756 & 0.708 & 0.634 & 0.774 & 0.660 \\
 & EMAF~\cite{EMAF} & 0.826 & 0.865 & 0.821 & 0.842 & 0.793 & 0.755 & 0.774 \\
 & MCAN~\cite{MCAN} & 0.809 & 0.889 & 0.765 & 0.822 & 0.732 & 0.871 & 0.795 \\
 & HMCAN~\cite{HMCAN} & 0.897 & \textbf{0.971} & 0.801 & 0.878 & 0.853 & 0.979 & 0.912 \\
 & CMMTN~\cite{CMMTN} & 0.903 & 0.870 & \textbf{0.917} & 0.892 & 0.927 & 0.881 & 0.903 \\
 & CCGN~\cite{CCGN} & 0.906 & 0.961 &0.748 & 0.841 & 0.886 &\textbf{0.984} & 0.933 \\
 & \textbf{KEN} & \textbf{0.934} & 0.912 & 0.911 & \textbf{0.911} & \textbf{0.947} & 0.948 & \textbf{0.947} 
 \\ \hline
\end{tabular}}
\end{center}
\label{tab:all}
\end{table*}

\subsection{Performance Comparison}
After extensive experiments, the results are presented in Table \ref{tab:all}. We find that our proposed KEN significantly outperforms all the considered baselines in terms of Accuracy and F1-score, which demonstrates the effectiveness and superiority of our model.

Although Att\_RNN and EANN integrate multimodal information, their performance is poor compared to other methods because they use LSTM and Text-CNN respectively to extract text features without leveraging BERT to obtain more powerful representations.
These observations indicate the importance of pre-trained language models in improving detection performance.
SAFE and CAFE learn the semantic consistency and similarity between different modalities, demonstrating the effectiveness of improving image-text alignment and learning consistency.
MKEMN and EMAF effectively extract news features and acquire knowledge and concepts from the entity perspective; however, insufficient semantic understanding and weak cross-modal alignment limit their fusion performance.
MCAN and HMCAN achieve better performance by introducing the co-attention mechanism, indicating the effectiveness of cross-modal semantic interaction and fine-grained fusion.
CMMTN leverages a mask-attention mechanism to learn intra- and inter-modal relationships, attenuating irrelevant features and reducing noise, while the integration of a curriculum-based PU learning method further enhances its detection performance.
CCGN improves image-text alignment through contrastive learning, and achieves superior performance by combining consistency learning with fine-grained cross-modal interaction fusion.
However, our proposed KEN model achieves the best performance.
We attribute the superiority of the KEN model to the following two reasons:

1) We obtain captions and evidence separately for images and text. Through interaction and complementation, we address previous research limitations in not fully understanding the scenes and semantics in images and the difficulty of effectively identifying false components when textual information is insufficient.

2) We fully explore the emotional features of news. By introducing a balanced learning strategy, we alleviate the performance degradation caused by previous research that processed all emotional types of news through a unified fake news detector.
Additionally, considering the tendency of fake news to express negative emotions, we design an auxiliary task to further improve detection performance.

{
\begin{table}[t]
\setlength{\abovecaptionskip}{0cm}
\caption{Ablation studies on different modalities.}
\centering
\resizebox{0.43\textwidth}{!}{
\begin{tabular}{ccccc}
\hline
\multirow{2}{*}{Dataset} & \multirow{2}{*}{Methods} & \multirow{2}{*}{Accuracy} & \multicolumn{2}{c}{F1-score} \\ \cline{4-5} 
 &  &  & Fake News & Real News \\ \hline
\multirow{5}{*}{Weibo} & Only Text & 0.905 & 0.906 & 0.904 \\
 & + Evidence & 0.919 & 0.921 & 0.917 \\ \cline{2-5} 
 & Only Image & 0.777 & 0.783 & 0.770 \\
 & + Caption & 0.790 & 0.790 & 0.789 \\ \cline{2-5} 
 & \textbf{KEN} & \textbf{0.935} & \textbf{0.935} & \textbf{0.934} \\ \hline
\multirow{5}{*}{Twitter} & Only Text & 0.830 & 0.745 & 0.873 \\
 & + Evidence & 0.843 & 0.771 & 0.881 \\ \cline{2-5} 
 & Only Image & 0.912 & 0.881 & 0.929 \\
 & + Caption & 0.919 & 0.892 & 0.936 \\ \cline{2-5} 
 & \textbf{KEN} & \textbf{0.934} & \textbf{0.911} & \textbf{0.947} \\ \hline
\end{tabular}
}
\label{ablation1}
\end{table}}

\subsection{Ablation Study}

\subsubsection{The Effect of Different Modalities}

We investigate the impact of different modalities using only text or image content.
The results are shown in Table \ref{ablation1}.
First, we find that the performance using only unimodal features is very poor, indicating the necessity of studying multimodal methods, as the interaction and fusion of information from different modalities help improve detection performance.
Secondly, we observe the differences between the two datasets. In Weibo dataset, the text modality conveys richer meanings and cues compared to the visual modality. In contrast, in Twitter dataset, the visual modality contains more content and information that helps detect authenticity.
Finally, by using the corresponding knowledge information on top of the unimodal features for interaction, supplementation, and fusion, the detection performance improves, providing preliminary validation of the effectiveness of our proposed knowledge-enhanced and interaction-fusion methods.

{
\begin{table}[t]
\setlength{\abovecaptionskip}{0cm}
\caption{Ablation studies on knowledge augmentation.}
\centering
\resizebox{0.43\textwidth}{!}{
\begin{tabular}{ccccc}
\hline
\multirow{2}{*}{DataSet} & \multirow{2}{*}{Methods} & \multirow{2}{*}{Accuracy} & \multicolumn{2}{c}{F1-score} \\ \cline{4-5} 
 &  &  & Fake News & Real News \\ \hline
\multirow{6}{*}{Weibo} & w/o KA & 0.915 & 0.916 & 0.913 \\
 & w/o CLIP & 0.911 & 0.911 & 0.911 \\
 & w/o Evidence & 0.917 & 0.918 & 0.917 \\
 & w/o Caption & 0.923 & 0.923 & 0.923 \\
 & w/o CA & 0.926 & 0.926 & 0.925 \\
 & \textbf{KEN} & \textbf{0.935} & \textbf{0.935} & \textbf{0.934} \\ \hline
\multirow{6}{*}{Twitter} & w/o KA & 0.912 & 0.884 & 0.929 \\
 & w/o CLIP & 0.889 & 0.845 & 0.914 \\
 & w/o Evidence & 0.914 & 0.883 & 0.932 \\
 & w/o Caption & 0.923 & 0.897 & 0.938 \\
 & w/o CA & 0.915 & 0.889 & 0.932 \\
 & \textbf{KEN} & \textbf{0.934} & \textbf{0.911} & \textbf{0.947} \\ \hline
\end{tabular}
}
\label{ablation2}
\end{table}}

\subsubsection{The Effect of Knowledge Augmentation}

We analyze the effectiveness of the knowledge augmentation method by removing relevant components.
The results are shown in Table \ref{ablation2}, where w/o KA removes the entire knowledge augmentation module, w/o CLIP removes CLIP-enhanced high-similarity image and text features, w/o Evidence removes external evidence supplementing text, w/o Caption removes caption information explaining image content and scenes, and w/o CA replaces the co-attention mechanism with concatenation.
We observe that:
1) The removal of the Knowledge Augmentation module leads to a significant performance degradation, demonstrating the soundness and effectiveness of our proposed approach to knowledge augmentation and interaction fusion;
2) The w/o CLIP has a greater impact than w/o KA because the semantic gap between cross-modal features in multimodal methods leads to a decrease in fusion performance.
CLIP alleviates this issue by mapping cross-modal features to the same semantic space through the image-text matching task. 
Furthermore, we observe a more significant performance decline on Twitter compared to Weibo, indicating that the latter has a higher degree of raw image-text similarity than the former;
3) The w/o Evidence shows that external evidence helps improve performance by breaking the information silos caused by closed and limited text and context.
The w/o Caption indicates  that mining image content and scenarios helps enhance the model's understanding of the news semantics and improves its discriminative ability;
4) The w/o CA reflects the semantic gap between image-text features and knowledge features. Direct concatenation introduces noise and lacks fusion and supplementation, leading to performance degradation, demonstrating the effectiveness of the co-attention mechanism in enhancing cross-modal fusion.

\subsubsection{The Effect of Emotion Guidance}

We analyze the effectiveness of the emotion guidance method by removing relevant components. 
The results are shown in Table \ref{ablation3}, where w/o EG removes all emotion-related components, including emotion analysis and balanced learning, w/o BL removes balanced learning implying the use of a unified detector for all emotional types of news, w/o Gate removes the gating mechanism replacing it with an averaging operation, and w/o ER removes the emotion reasoning task. 
We observe that:
1) The w/o EG shows that removing all emotion-related components leads to a significant performance degradation, indicating that emotional features contribute to fake news detection and further demonstrating the effectiveness of our proposed emotion guided method.
2) The introduction of balanced learning improves the model’s detection performance because our approach allows news detectors of different emotional types to analyze the content they specialize in, avoiding the performance degradation caused by previous approaches that give all emotional types of news to a uniform detector.
3) The w/o Gate indicates that the attention weights obtained through the gating mechanism effectively allow the detectors corresponding to the news emotional  types to be prominently expressed.
4) The w/o ER demonstrates the effectiveness of considering the phenomenon that fake news tends to express negative emotions in improving the model's detection performance.

{
\begin{table}[t]
\setlength{\abovecaptionskip}{0cm}
\caption{Ablation studies on emotion guidance.}
\centering
\resizebox{0.43\textwidth}{!}{
\begin{tabular}{ccccc}
\hline
\multirow{2}{*}{DataSet} & \multirow{2}{*}{Methods} & \multirow{2}{*}{Accuracy} & \multicolumn{2}{c}{F1-score} \\ \cline{4-5} 
 &  &  & Fake News & Real News \\ \hline
\multirow{5}{*}{Weibo} & w/o EG & 0.918 & 0.918 & 0.918 \\
 & w/o BL & 0.922 & 0.923 & 0.921 \\
 & w/o Gate & 0.926 & 0.928 & 0.925 \\
 & w/o ER & 0.929 & 0.930 & 0.928 \\
 & \textbf{KEN} & \textbf{0.935} & \textbf{0.935} & \textbf{0.934} \\ \hline
\multirow{5}{*}{Twitter} & w/o EG & 0.912 & 0.884 & 0.928 \\
 & w/o BL & 0.916 & 0.889 & 0.932 \\
 & w/o Gate & 0.919 & 0.891 & 0.935 \\
 & w/o ER & 0.923 & 0.898 & 0.938 \\
 & \textbf{KEN} & \textbf{0.934} & \textbf{0.911} & \textbf{0.947} \\ \hline
\end{tabular}
}
\label{ablation3}
\end{table}}

\subsection{Parameter Analysis}

\subsubsection{Impact of the Number of Experts}

\begin{figure}[t]
\centering
\scalebox{0.95}
{\includegraphics[width=1\linewidth]{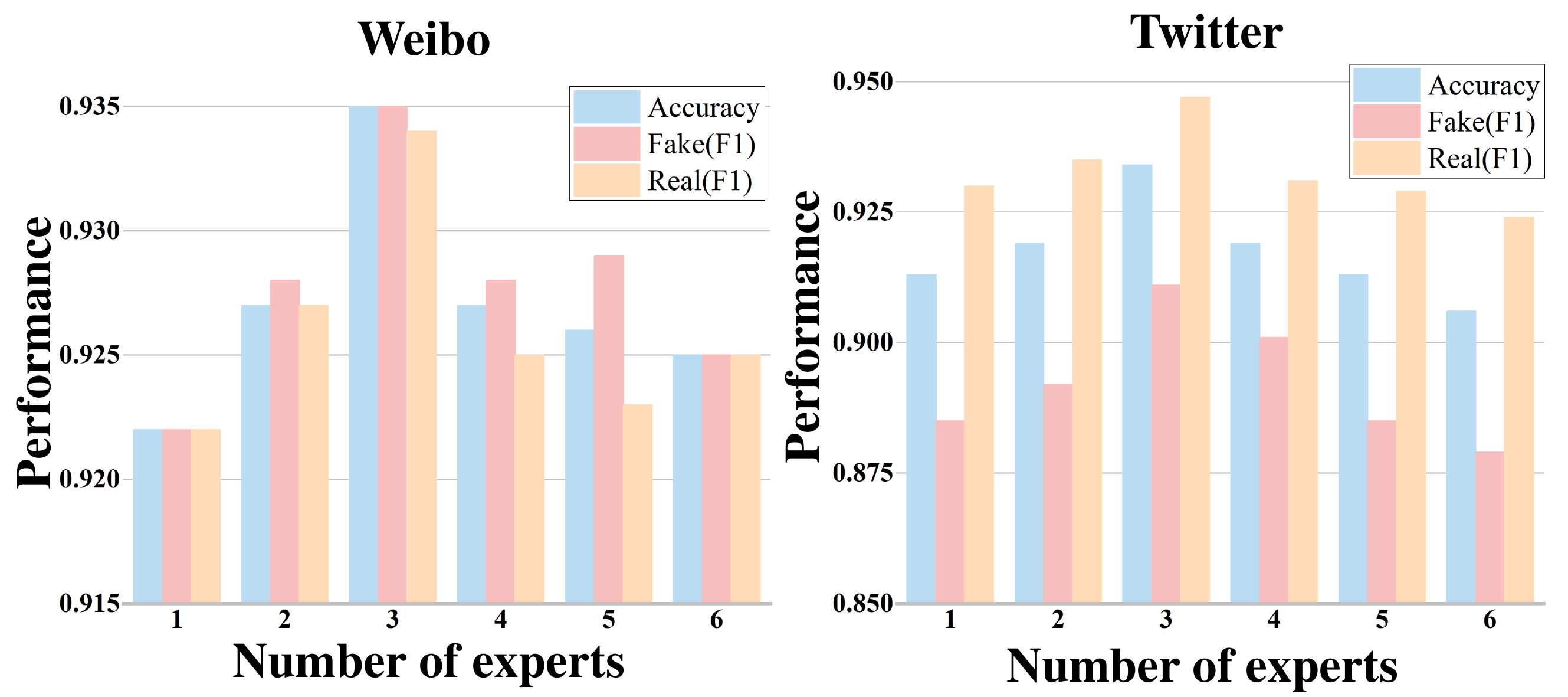}}
\caption{
Impact of different number of experts ($k$) on detection performance.}
\label{fig:p1}
\end{figure}

In emotion analysis, we set up multiple experts to collaboratively learn emotion features.
We conducted experiments on the Weibo and Twitter datasets using varying numbers of experts ($k$), and the results are shown in Figure~\ref{fig:p1}. 
The experimental results indicate that the detection performance exhibits a trend of first increasing and then decreasing as $k$ changes. When $k=3$, the KEN achieves the best performance on both datasets.
When the number of experts is small, the opinions of a single expert may be highly subjective or biased. Such extreme emotion analysis can adversely affect the stability and generalization ability of the model, leading to lower detection performance.
On the other hand, when the number of experts is too large, their opinions may contain redundant information and conflicting viewpoints, making it difficult for the model to effectively analyze emotion features, which also results in decreased performance.
Ultimately, an appropriate number of experts ($k=3$) strikes a balance between diversity and consistency, enabling the model to capture rich emotion features while avoiding the noise interference caused by excessive information.

\setlength{\abovecaptionskip}{0cm}

\setlength{\abovecaptionskip}{0cm}
\begin{figure}[t]
\centering
\scalebox{0.95}
{\includegraphics[width=1\linewidth]{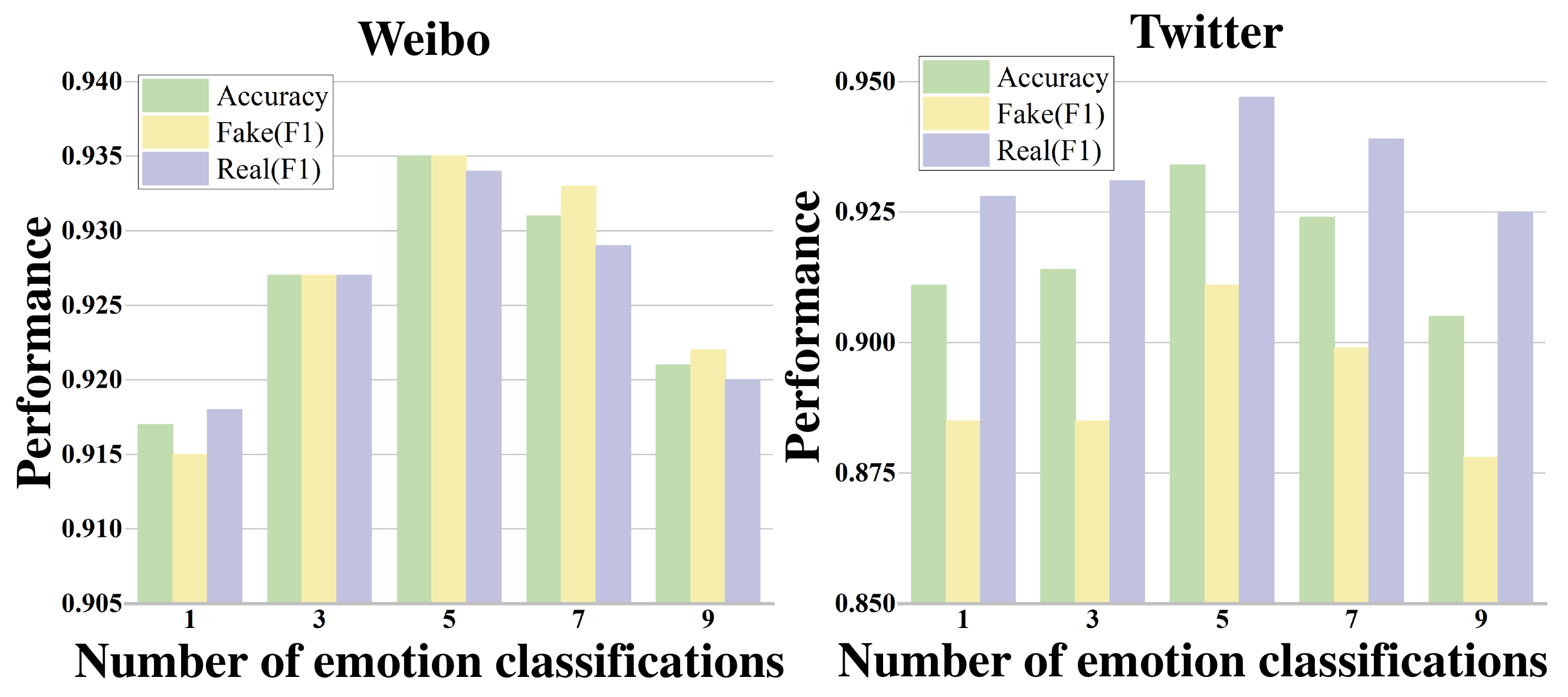}}
\caption{
Impact of different number of emotional  types ($x$) on detection performance.}
\label{fig:p2}
\end{figure}

\subsubsection{Impact of the Number of Emotional Types}

In balanced learning, we emphasize processors corresponding to the emotional  types of news, effectively addressing the performance degradation caused by feeding different emotional  types of news into a unified processor in previous studies.
We conducted experiments on the Weibo and Twitter datasets using varying numbers of emotional  types ($x$), and the results are shown in Figure~\ref{fig:p2}. 
We observe that KEN achieves optimal performance on both datasets when $x=5$.
When the number of emotional  types is small, the model fails to adequately distinguish features of different emotion categories, resulting in the neglect of some fine-grained emotional information. In such cases, certain emotional types may be roughly or incorrectly classified into other categories, causing the detector to fail to capture their unique emotional features, ultimately leading to a decline in detection performance.
When the number of emotional  types is large, the complexity of classification increases, potentially introducing more noise and the risk of misclassification. Additionally, the boundaries between emotions may become blurred, making it difficult to highlight processors corresponding to news emotional  types and potentially leading to overfitting.

\subsubsection{Emotional Importance from Different Modalities}
\begin{figure}[t]
\centering
\scalebox{0.95}
{\includegraphics[width=1\linewidth]{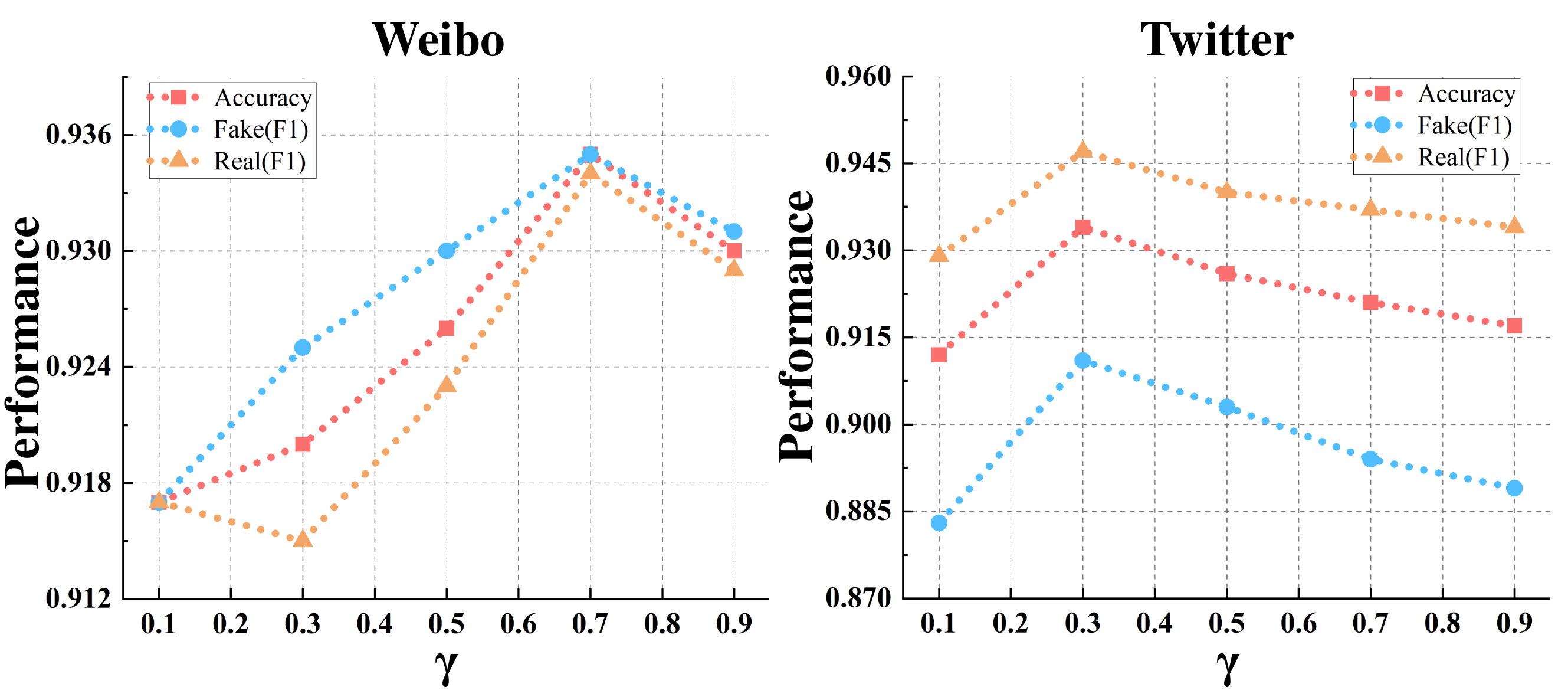}}
\caption{
Impact of emotion importance adjustment between text and image modalities.}
\label{fig:p3}
\vspace{-0.4cm}
\end{figure}

In emotion analysis, we capture overall news emotion features by integrating text and image emotions, using a modulation parameter $\gamma$ (0 to 1) to balance their importance,
where a larger value indicates the text emotion is more important, while a smaller value highlights the image emotion.
We conducted experiments on the Weibo and Twitter datasets with various values of $\gamma$, and the results are shown in Figure~\ref{fig:p3}.
We observe that the overall performance of KEN on both datasets shows an initial increase followed by a decrease, indicating the presence of an optimal solution. 
In the Weibo dataset, the model achieves the best performance when $\gamma$ is set to 0.7, suggesting that the text modality conveys richer and more critical emotional information than the image modality. 
In contrast, in the Twitter dataset, the model achieves optimal performance when $\gamma$ is set to 0.3, indicating that the image modality contains more and more significant emotional cues than the text modality.

\setlength{\abovecaptionskip}{0cm}

\setlength{\abovecaptionskip}{0cm}
\begin{figure}[t]
\centering
\scalebox{0.95}
{\includegraphics[width=1\linewidth]{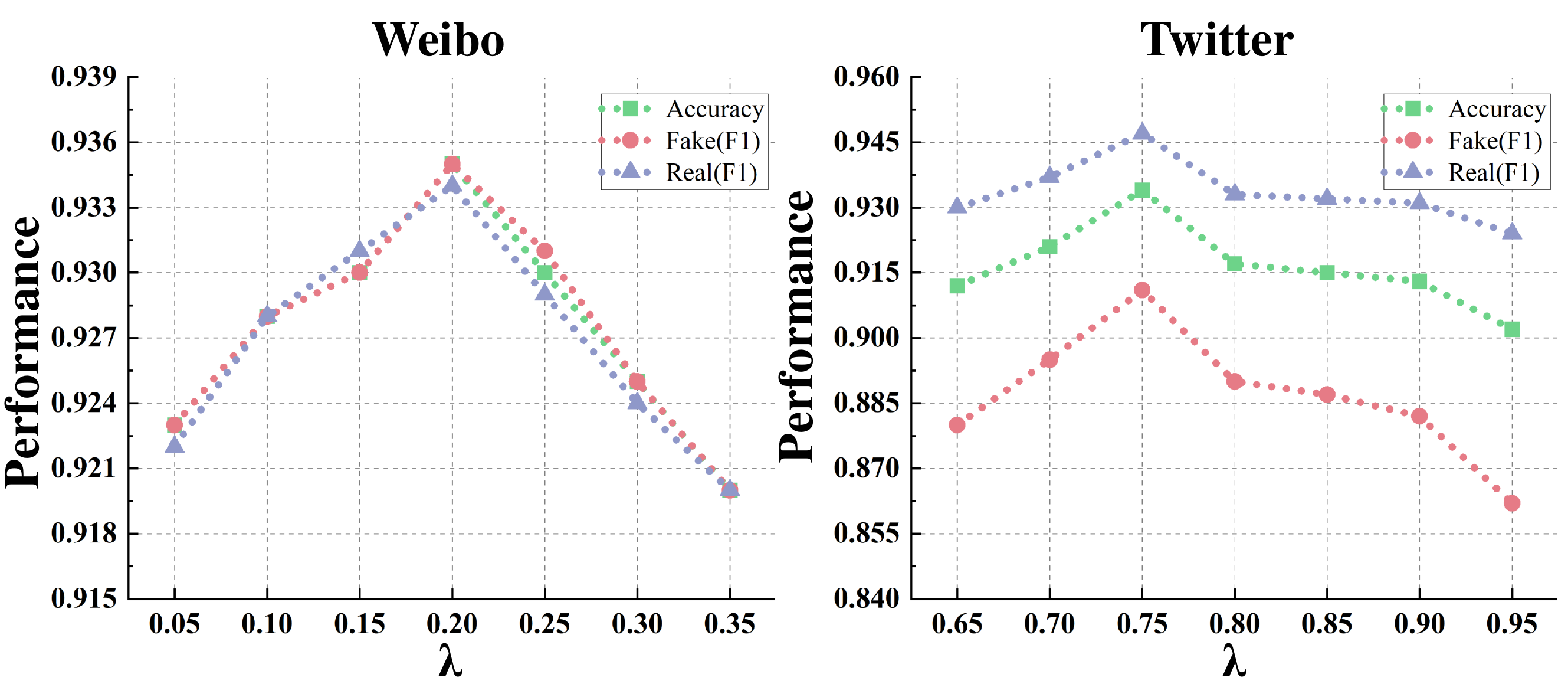}}
\caption{
Impact of adjusting the loss balance parameter $\lambda$.}
\label{fig:p4}
\end{figure}

\subsubsection{Impact of the Loss Balance Parameter $\lambda$}
Considering that fake news in the real world tends to express negative emotions, we design an auxiliary task to leverage this phenomenon. 
We roughly assume that news with negative emotions is fake, while news with positive emotions is real, thus constructing an emotion reasoning loss.
The overall model loss consists of the emotion reasoning loss from the auxiliary task and the fake news detection loss from the main task, with the relationship between the two being adjusted by the loss balance parameter $\lambda$. 
We conducted experiments with different values of $\lambda$ on the Weibo and Twitter datasets, and the results are shown in Figure~\ref{fig:p4}. 
We observe that when $\lambda$ is set to 0.20 on the Weibo dataset, the model achieves the best performance, while on the Twitter dataset, the model performs best when $\lambda$ is set to 0.75. The experimental results indicate that emotional features in Twitter play a more significant role in inferring the authenticity of news compared to Weibo, and the correlation between emotional tendency and news authenticity is stronger on Twitter than on Weibo. 
This may be because news data on Twitter is often presented as direct user comments on events, while news on Weibo is more like formal reports and reposts.
The emotions in the former are more intense, and the correlation between news authenticity and emotional tendency is more pronounced.

\subsection{Visualization Results}

\begin{figure}[t]
\centering
\scalebox{0.95}
{\includegraphics[width=1\linewidth]{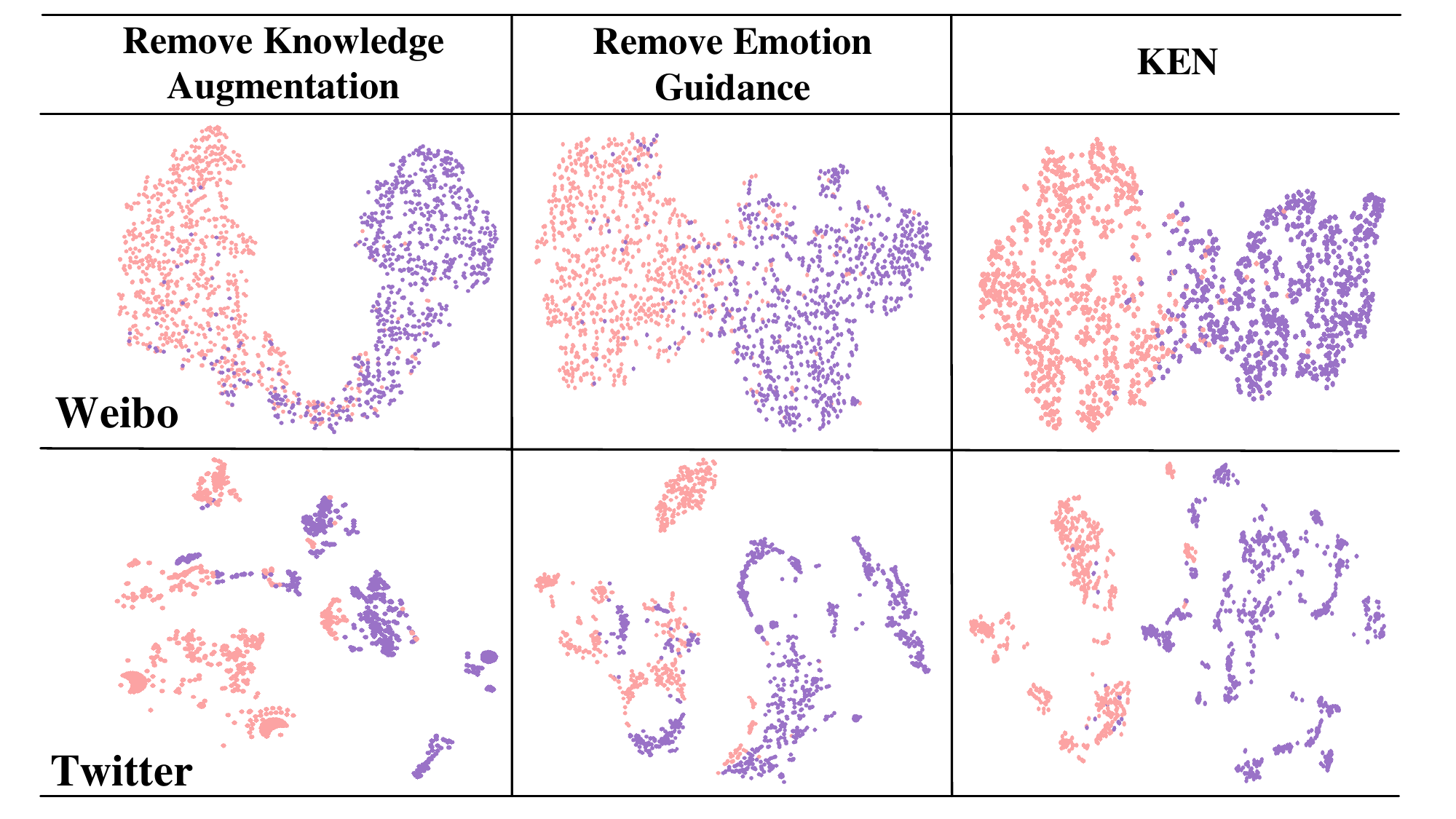}}
\caption{
T-SNE visualization of mined features on the test sets of Weibo and Twitter.}
\label{fig:tsne}
\end{figure}

To intuitively demonstrate the effectiveness of the proposed knowledge augmentation and emotion guidance methods, we applied t-SNE~\cite{tsne} to visualize the classification features after removing the relevant modules of KEN on the test sets of Weibo and Twitter.
The results are shown in Figure~\ref{fig:tsne}, where points of the same color represent the same authenticity label.
We observed that the aggregated features of Twitter are more dispersed and form multiple clusters compared to Weibo. 
This is because Weibo dataset is more balanced, whereas Twitter consists of news centered around several major events. 
Therefore, news from the same event have closer features, while news from different events have greater distance.
Additionally, we can clearly see that the complete KEN model has smaller overlapping regions and is able to more clearly distinguish between real and fake news, achieving the best classification performance. 
This fully demonstrates the effectiveness of the proposed knowledge augmentation and emotion guidance methods.

\setlength{\abovecaptionskip}{0cm}

\section{Conclusion}
In this work, we propose a novel \textbf{K}nowledge Augmentation and \textbf{E}motion Guidance \textbf{N}etwork (KEN) for multimodal fake news detection.
On the one hand, we effectively leverage LVLM’s powerful semantic understanding and extensive world knowledge.
For the visual modality, the generated captions provide a comprehensive understanding of image content and scenes, while for the textual modality, the retrieved evidence helps break the information silos caused by the closed and limited text and context. 
By designing three fusion perspectives and integrating the co-attention mechanism with CLIP, we further enhance the semantic representation and fusion performance of news.
On the other hand, we consider inter-class differences among news with different emotions. Through balanced learning, we model the fine-grained relationship between emotional types and authenticity, avoiding performance degradation caused by treating all emotional types of news uniformly in traditional methods.
We conduct comprehensive comparative and ablation experiments on two real-world datasets, demonstrating the effectiveness and superiority of our proposed methods.

\begin{acks}
This work was supported in part by the National Natural Science Foundation of China (62472117), the Guangdong Basic and Applied Basic Research Foundation (2025A1515010157), the Science and Technology Projects in Guangzhou (2025A03J0137).
\end{acks}

\bibliographystyle{ACM-Reference-Format}
\bibliography{sample-sigconf}

\end{document}